# Gated Recurrent Unit based Autoencoder for Optical Link Fault Diagnosis in Passive Optical Networks

Khouloud Abdelli [(1,2)], Florian Azendorf [(3)], Helmut Grießer [(1)], Carsten Tropschug [(3)], Stephan Pachnicke [(2)]

[(1)] ADVA Optical Networking SE, Fraunhoferstr. 9a, 82152 Munich, Germany, kabdelli@adva.com
[(2)] Christian-Albrechts-Universität zu Kiel, Kaiserstr. 2, 24143 Kiel, Germany
[(3)] ADVA Optical Networking SE, Märzenquelle 1-3, 98617 Meiningen, Germany

**Abstract** We propose a deep learning approach based on an autoencoder for identifying and localizing fiber faults in passive optical networks. The experimental results show that the proposed method detects faults with 97% accuracy, pinpoints them with an RMSE of 0.18 m and outperforms conventional techniques.

## Introduction

Passive Optical Networks (PONs) have been emerging as an important broadband access network technology. Due to the growing transmission capacity of PONs, it is getting more and more important to ensure their survivability and reliability. To automatically monitor the PON fiber infrastructure helps to quickly detect and pinpoint potential faults, thereby improving availability and saving costs. The optical link monitoring in PONs has been mainly carried out using the optical time domain reflectometry (OTDR) technique widely used for characterizing an optical link and for fiber fault detection and localization. Even if a change in the reflected signal can indicate a potential problem in the PON system, it still might be difficult to detect the type of error and to precisely identify the root cause. To make it easier to monitor the loss of a branch and to demarcate the network, often reference reflectors are placed at the termination point of each branch[1], but this does require changes in the PON network and does not apply to all types of faults. Lee[2] proposed to leverage the wavelength dependence of certain faults to help identifying them, for this the OTDR needs to perform its measurements at different wavelengths adding to hardware complexity and measurement time. Without relying on specific hardware requirements, a DSP based approach can help to classify events, e.g. Kong[3] could distinguish reflective and non-reflective faults by employing correlation matching and a short time Fourier transform.

Recently, OTDR event analysis using machine learning (ML) techniques for fault detection and localization has been demonstrated[4-6], with the last publication also differentiating between reflective and non-reflective events.

In this paper, we go one step further and present a novel gated recurrent unit (GRU) based autoencoder model, called GRE-AE in the following, that automatically identifies a broad range of fiber optical faults in PONs and fully characterizes it without requiring either the intervention of trained personnel or specific hardware. To the best of our knowledge, this is the first time that the classification of the fiber faults and particularly the fiber bending event is investigated without involving any analysis of the wavelength dependence of the loss. The GRU-AE approach is applied to experimental OTDR data incorporating different fiber faults such as fiber tapping, link breaks and fiber bendings. The results show that our model: (i) diagnoses the faults with a high accuracy and locates them with a minimal error and within short measurement time; and (ii) outperforms a conventional method.

## Experimental Setup and Data Generation

The experimental setup shown in Fig. 1 is carried out to record OTDR traces comprising various faults in PON network. To reproduce a real PON envionment, 4 splitters are utilized. Various faults (i.e. events) namely *fiber tapping*, *attenuation splice*, *dirty connector*, *fiber bending*, *fiber break, PC connector* and *reflector* are induced at different locations in the network. For the fiber bending and tapping events, bend radius values of 10 mm, 7.5 mm, 5 mm, 2.5 mm are applied to generate faults with different losses (i.e. different event patterns). Different bad attenuation splices with dissimilar losses are conducted to create a varying bad splicing fault pattern. Reflectors, PC and open PC connectors cause reflective events with various reflectances. To adjust the height of the reflective peak due to the reflector, a variable optical attenuator (VOA) is used that allows to change the attenuation from 0 to 30 dB. The OTDR configuration parameters namely the pulse width, the wavelength and the sampling time are set to 10 ns, 1650 nm and 1 ns, respectively. For every OTDR record collected, from 62 up to 65,000 traces are averaged to reduce the noise of the signal.

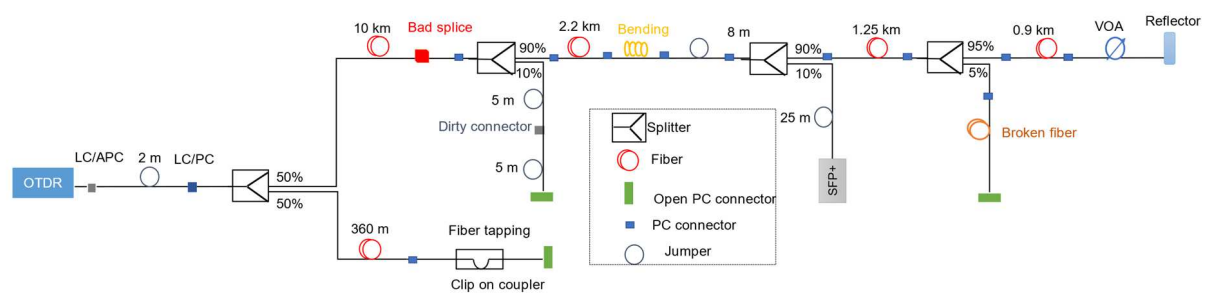

**Fig. 1:** Experimental setup for generating OTDR data incorporating several faults at different locations in a PON (PC - physical contact, APC – angled PC, LC – Lucent connector)

**Data Preprocessing**

The recorded OTDR traces are split into sequences of length 30. For each sequence, the event type (*no event*, or one of the aforementioned events), the event position index within the sequence, and the event reflectance and/or loss are assigned. Since the SNR significantly impacts the event pattern and the amplitude of the signal is an important characteristic of the reflective event, the SNR ($\gamma$) and a feature $\delta$ defined as the maximum of the signal amplitude are estimated for each sequence. To balance the event type distribution in the data used to train the ML model, an equal number of samples (instances) for each event is randomly selected. In total, a data set composed of 125,752 samples, whose SNR values vary from 0 to 30 dB, is built. The data is normalized and randomly divided into training, validation and test datasets with a ratio of 60/20/20.

**GRU-AE Model**

The overall architecture of the GRU-AE model is depicted in Fig. 2. The input of the model consists of a sequence of power levels (of length 30) combined with $\delta$ and $\gamma$. Fed with the input, the shared GRU based encoder is applied to extract the relevant features modelling the event pattern. It comprises two GRU layers containing 30 and 15 cells, respectively. The GRU is chosen as it is well-suited to process sequential data and to adaptively capture dependencies of the data due to its gates (i.e. the update and the reset gate) controlling the flow of the information. The learned representation (extracted features) by the shared encoder is transferred to several decoders for performing the fault diagnosis, event localization and characterization. Each decoder consists of two GRU layers with 15 and 30 cells followed by a fully connected layer composed of 16 neurons. The overall loss of the model is computed as the weighted sum of the four individual decoder losses set to 1, 1.5, 1 and 1, respectively. The optimizer chosen is the Adam optimization algorithm.

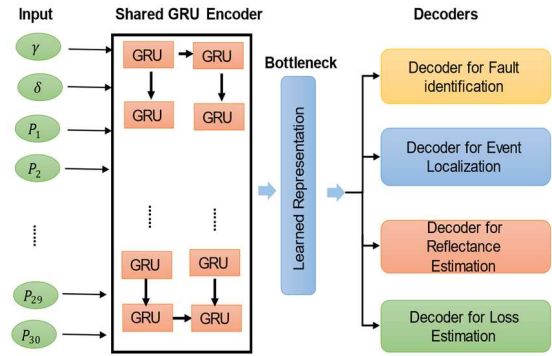

**Fig. 2:** Proposed GRU-AE architecture composed of an encoder transferring the learned knowledge to different decoders for fault diagnosis, localization, and characterization

**Results and Discussion**

To evaluate the event detection capability of the GRU-AE model, several metrics including the confusion matrix showing the misclassification rates, the detection probability ($P_d$) and the false alarm rate ($P_{FA}$) are used.

|  | No event | Tapping | Bad splice | Bending | Dirty connector | Broken fiber | Reflector | PC connector |
|---|---|---|---|---|---|---|---|---|
| PC connector |  |  |  |  | 0.004 |  | 0.006 | 0.989 |
| Reflector |  | 0.002 | 0.001 | 0.001 | 0.022 |  | 0.967 | 0.006 |
| Broken fiber |  |  | 0.001 | 0.065 |  | 0.932 | 0.002 |  |
| Dirty connector |  | 0.001 |  |  | 0.983 |  | 0.012 | 0.004 |
| Bending |  |  | 0.001 | 0.930 |  | 0.067 | 0.001 |  |
| Bad splice |  |  | 0.997 | 0.001 |  |  | 0.001 |  |
| Tapping | 0.004 | 0.994 |  |  |  |  | 0.001 |  |
| No event | 0.997 | 0.002 |  |  |  |  | 0.001 |  |

**Fig. 3:** Confusion matrix of the GRU-AE model

As shown in Fig. 3, the GRU-AE model detects the different faults with an accuracy higher than 93%. For the bending and broken fiber events, the accuracy is lower compared to the other faults as the event patterns of the aforementioned faults are very similar particularly for low SNR sequences, and thereby the GRU-AE model misclassified these events. The PC connector is misclassified rarely as either reflector or dirty connector due to the similarity between these reflective events. The GRU-AE model misclassified the reflector sometimes as either dirty connector or open PC connector especially for high SNR sequences due to the high reflection observed for these events. For sequences with low SNR values and due to the high attenuation set by the VOA, the height of the reflection is reduced significanly. As an attenuation is observed after the reflective peak, the pattern of the event is similar to the non-reflective events for such cases. Therefore, the GRU-AE model missclassified the reflector as bending, tapping or bad splice events.

As depicted in Fig. 4, the detection probability $P_d$ of the different events increases with SNR, and it is approaching 1 for SNR values higher than 10 dB. For lower SNR values ($SNR \leq 10$ dB), $P_d$ is lower mainly due to noise that influences the event pattern and makes it look similar to random noise spikes (no event pattern). Thereby it is tricky for the GRU-AE model to distinguish between the events.

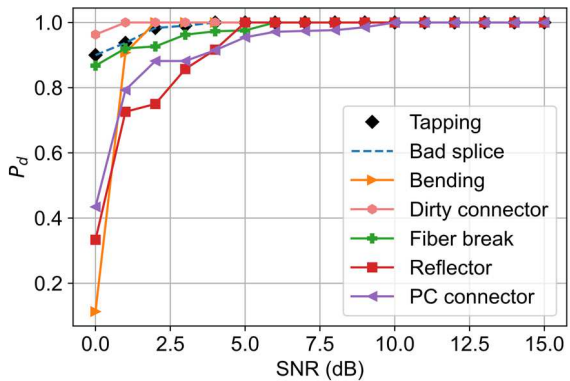

**Fig. 4:** Detection probability of the different faults as function of SNR

As shown in Fig. 5, the false alarm rate $P_{FA}$ for the different faults decreases with SNR and it is generally low, but the relatively high value at 0 dB for the dirty connector case indicates that the detection probability for this is too optimistic.

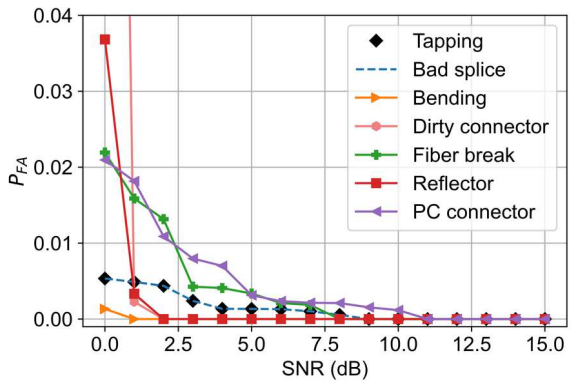

**Fig. 5:** False alarm rate of the different faults as function of SNR, the false alarm rate of the dirty connector at SNR 0 dB is 0.25

The capability of the GRU-AE model in locating the event and characterizing it in terms of reflectance and/or loss as function of SNR is also investigated. As shown in Fig. 6, the event position error is lower than 0.33 m and can be less than 0.15 m for high SNR values. The reflectance estimation error is less than 3.5 dB, and it can reach up to 0.5 dB for high SNR values, whereas for the loss prediction error, it is less than 1.4 dB. For high SNR values it can even be reduced up to 0.2 dB.

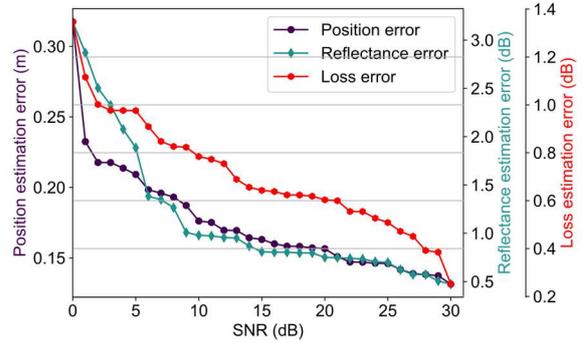

**Fig. 6:** Event position, reflectance, and loss estimation error for the GRU-AE model

The GRU-AE approach is compared to a conventional OTDR event analysis method based on combining the matched template incorporating the typical reflective and non-reflective event patterns and setting event thresholds, using an unseen test dataset slightly different from the training dataset (different laser powers, averaging and SNR values, modified bending radius of the tapping and bending events, different attenuation, slightly changed PON network (one of the splitters is omitted) and thereby varied position of the reflector event). Given that the conventional method detects the event without being able to identify the type of the fault, we just compare the capability of both approaches in detecting the fault regardless of the event type. The results shown in Fig. 7 demonstrate that the GRU-AE model outperforms the conventional technique by achieving higher accuracy and smaller root mean square error (RMSE).

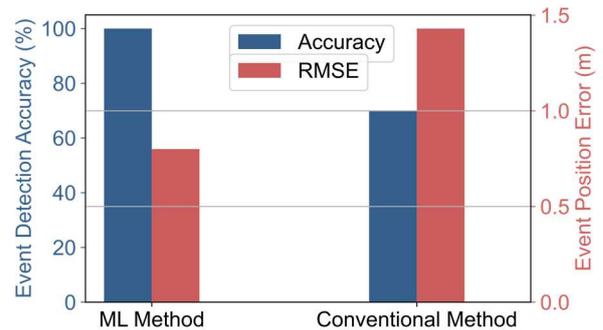

**Fig. 7:** Comparison of the GRU-AE approach and conventional method on unseen test dataset using the fault accuracy and the event position error metrics

**Conclusions**

In this paper, we presented a GRU-based auto-encoder model for automatic fiber fault identification and characterization in PONs. The results demonstrated that the approach achieves a great detection and localization accuracy and outperforms a conventional method.